\documentstyle[12pt,epsfig]{article}
\advance\hoffset by -7mm  
\makeatletter
\@addtoreset{equation}{section}
\makeatother

\renewcommand{\title}[1]{\null\vspace{25mm}

\noindent{\Large{\bf #1}}\vspace{10mm}

\noindent {\large By }}
\newcommand{\authors}[1]{\noindent{\large #1}\vspace{3mm}

}
\newcommand{\address}[1]{\noindent #1\vspace{5mm}

}
\renewcommand{\abstract}[1]{\vspace{19mm}

\noindent{\small{\em Abstract.} #1}\vspace{2mm}

} 
\newcommand{\be}{\begin{equation}}
\newcommand{\ee}{\end{equation}}

\begin{document}
\title{Early History of Gauge Theories and Weak \\
Interactions\footnote{Invited talk at the \underline{PSI Summer School on
Physics with Neutrinos}, Zuoz, Switzerland, August 4-10, 1996.}}
\authors{Norbert Straumann}
\address{Institut f\"ur Theoretische Physik der
Universit\"at Z\"urich--Irchel,\\
Winterthurerstrasse 190, CH--8057 Z\"urich,
Switzerland}

\section{Introduction}

It took decades until physicists understood that all known
fundamental interactions can be described in terms of gauge theories.
My historical account begins with Einstein's general theory of
relativity (GR), which is a non-Abelian gauge theory of a special 
type (see Secs. 3,4). The other gauge theories emerged in a slow
 and complicated process gradually from GR, and their common geometrical
structure --- best expressed in terms of connections of fiber bundles
--- is now widely recognized. Thus, also in this respect, 
H. Weyl was right when he wrote in the preface to the first edition 
of Space -- Time -- Matter (RZM) early in 1918: 
``Wider expanses and greater depths are now exposed to the 
searching eye of knowledge, regions of  which  we  had  not  even  a 
presentiment. 
It has brought us much nearer to grasping the plan that
underlies all physical happening'' \cite{1}. 

It was Weyl himself who made in 1918 the first attempt to extend GR in order 
to describe gravitation and electromagnetism within a unifying
geometrical framework \cite{2}. This brilliant proposal contains all 
mathematical aspects of a non-Abelian gauge theory,
as I will make clear in $\S 2$. The words gauge (Eich--) transformation
and gauge invariance appear the first time in this paper, but in the 
everyday meaning of change of length or change of calibration.

Einstein admired Weyl's theory as  ``a coup of genius of the first rate 
\ldots '',
but immediately realized that it was physically untenable: ``Although your 
idea is so 
beautiful, I have to declare frankly that, in my opinion, it is impossible 
that 
the theory corresponds to nature.'' This led to an intense exchange of 
letters
between Einstein (in Berlin) and Weyl (at the ETH in Z\"urich), which will
hopefully soon be published in {\sl The Collected Papers} of Einstein.
(In my article \cite{3} I gave an account of this correspondence which 
is preserved in the Archives of the ETH.) No agreement was reached, but
Einstein's intuition proved to be right.

Although Weyl's attempt was a failure as a physical theory it paved the way
for the correct understanding of gauge invariance. Weyl himself re-interpreted
his original theory after the advent of quantum theory in a seminal paper 
\cite{4}
which I will discuss at length in $\S 3$. Parallel developments by other
workers
and interconnections are indicated in Fig.1. 

At the time Weyl's contributions to theoretical physics were not appreciated 
very 
much, since they did not really add new physics. The attitude of the leading
theoreticians is expressed in familiar distinctness in a letter by Pauli to 
Weyl
from July 1, 1929, after he had seen a preliminary account of Weyl's work:
\begin{quotation}
%\item

{\sl Before me lies the April edition of the Proc.Nat.Acad. (US). 
Not only does it contain an article from you under ``Physics''
but shows that you are now in a `Physical Laboratory': from what I hear 
you have even been given a chair in `Physics' in America. 
I admire your courage; since the conclusion is inevitable that you 
wish to be judged, not for success in pure mathematics, but for your
true but unhappy love for physics \cite{5}.}
\end{quotation}

Weyl's reinterpretation of his earlier speculative proposal had actually 
been suggested 
before by London, but it was Weyl who emphasized the role of gauge 
invariance as a {\em symmetry principle} from which electromagnetism 
can be {\em derived}. It took several decades until the importance 
of this symmetry principle --- in its generalized form to non-Abelian
gauge groups developed by Yang, Mills, and others --- became also 
fruitful for a description of the weak and strong interactions. 
The mathematics of the non-Abelian generalization of Weyl's 1929 paper
would have been an easy task for a mathematician of his rank, but at the time
there was no motivation for this from the physics side. The known properties
of the weak and strong nuclear interactions, in particular their short range,
did not point to a gauge theoretical description. We all know that the gauge
symmetries of the Standard Model are very hidden and it is, therefore,
not astonishing that progress was very slow indeed.

Today, the younger generation, who learned the Standard Model from
polished textbook presentations, complains with good reasons about many 
of its imperfections. It is one of the aims of this talk to make it obvious
that it was extremely difficult to reach our present understanding 
of the fundamental interactions. The Standard Model, 
with all its success, is a great achievement, and one should not be too
discouraged when major further progress is not coming rapidly. 

Because of limitations of time and personal knowledge, I will discuss in the 
rest of my talk mainly the two important papers by Weyl from 1918 and 1929. 
The latter contains also his two-component theory of massless spin $1/2$ fermions.
In this context I will make in $\S 5$ a few remarks about the developments
which led in 1958 to the phenomenological $V-A$ current-current 
Lagrangian for the weak interactions. My historical account of the 
non-Abelian generalizations by Klein, Pauli and others, culminating in the 
paper
by Yang and Mills, will also be much abbreviated. This is not too bad,
since there will soon be a book by Lochlain O'Raifeartaigh
that is devoted entirely to the early history of gauge theories \cite{6}. 
Those who do not know German will find there also English translations 
of the most important papers of the first period (1918--1929). 
The book contains in addition the astonishing paper by Klein (1938) \cite{7},
Pauli's letters to Pais on non-Abelian Kaluza-Klein reductions \cite{8},
parts of Shaw's dissertation, in which he develops a non-Abelian
$SU(2)$ gauge theory \cite{9}, and Utiyama's generalization of Yang-Mills
theory to arbitrary gauge groups \cite{10}. These works are behind the 
diagram in Fig.1.

This talk covers mostly material contained in the papers \cite{3}, 
\cite{11}, and
\cite{12}, which I have published some time ago in German, partly because 
all early publications and letters related to our subject
are written in this language. 

\begin{figure}
\begin{center}
\epsfig{file=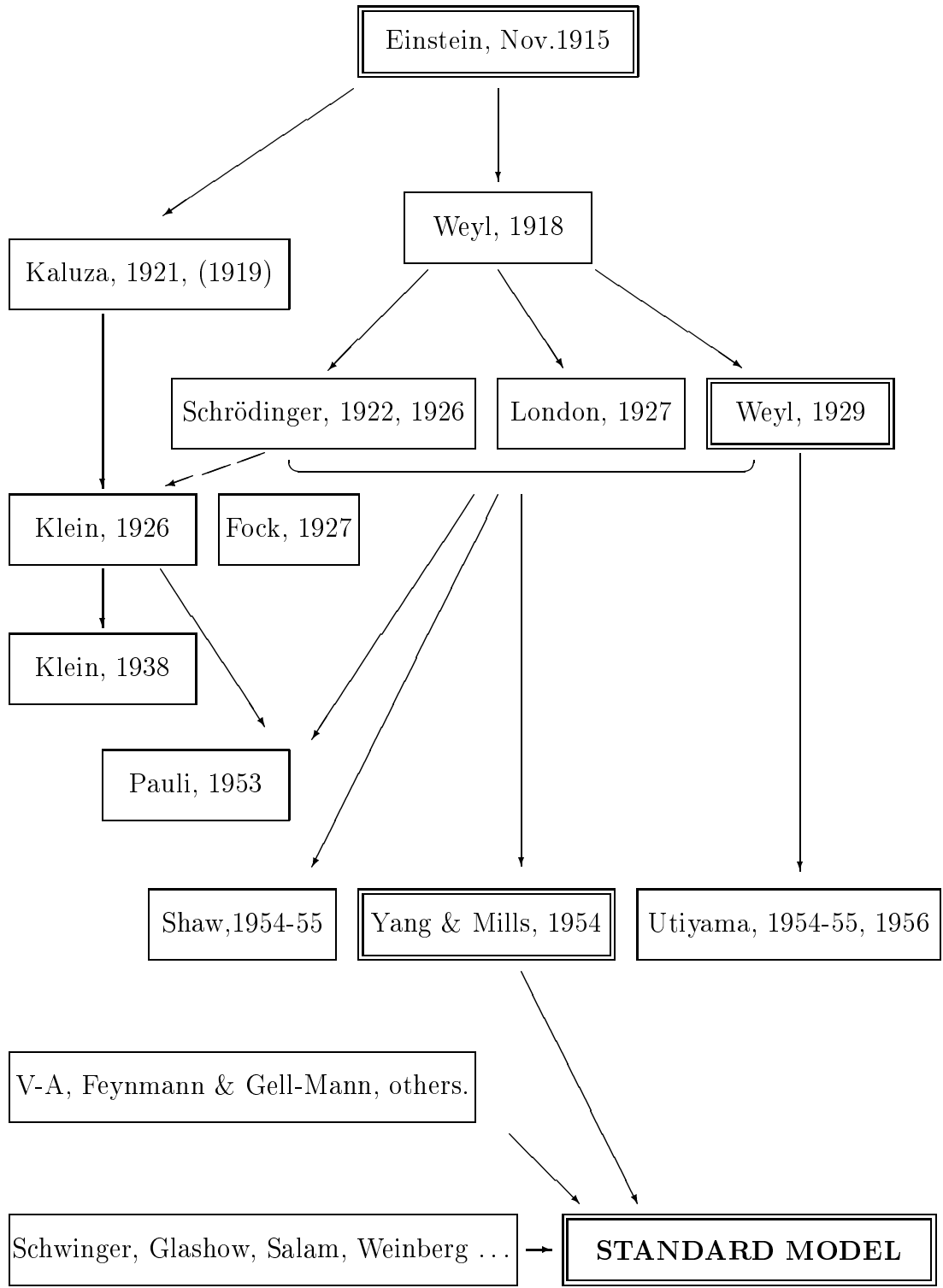,height=20cm}
%\vspace{-4cm}
\caption{Key papers in the development of gauge theories.}
\end{center}
\end{figure}

%\begin{center}
%{\bf Fig.1.} Key papers in the development of gauge theories.
%\end{center}
\clearpage

\section{Weyl's attempt to unify gravitation \newline
 and electromagnetism}

On the 1st of March 1918 Weyl writes in a letter to Einstein:
``These days I succeeded, as I believe, to derive electricity and gravitation
from a common source \ldots ''. Einstein's prompt reaction by postcard
indicates already a physical objection which he explained in detail shortly
afterwards. Before I come to this I have to describe Weyl's theory of 1918.

\subsection{Weyl's generalization of Riemannian geometry}

Weyl's starting point was purely mathematical. He felt a certain uneasiness
about Riemannian geometry, as is clearly expressed by the following sentences 
early
in his paper\footnote{I am using here and at other places the English
translation of L. O'Raifeartaigh \cite{6}.}:
\begin{quotation}
%\item
{\sl But in Riemannian geometry described above there is contained a last
element of geometry ``at a distance''  (ferngeometrisches Element) --- with no good
reason, as far as I can see; it is due only to the accidental development 
of Riemannian geometry from Euclidean geometry. The metric allows the
two magnitudes of two vectors to be compared, not only at the same point, but 
at any
arbitrarily separated points.} {\it A true infinitesimal geometry should, 
however,
recognize only a principle for transferring the magnitude of a vector to an
 infinitesimally close point} {\sl and then, on transfer to an arbitrary 
distant point, the integrability of the magnitude of a vector is no more to be 
expected that the
integrability of its direction.}
\end{quotation}

After these remarks Weyl turns to physical speculation and continues as follows:

\begin{quotation}
%\item
{\sl On the removal of this inconsistency there appears a geometry that,
surprisingly,
when applied to the world,} {\it explains not only the gravitational
phenomena but 
also the electrical.} {\sl According to the resultant theory 
both spring from the same source, indeed} {\it in general one 
cannot separate gravitation and electromagnetism in a unique manner}.
{\sl In this theory} {\it all physical
quantities have a world geometrical meaning; the action appears from the 
beginning as a pure number. It leads to an essentially unique universal law;
it even allows us to understand in a certain sense why the world is
four-dimensional}.
\end{quotation}

In brief, Weyl's geometry can be described as follows. First, the spacetime 
manifold $M$ is equipped with a conformal structure, i.e.,
with a class $[g]$ of
conformally equivalent Lorentz metrics $g$ (and not a
definite metric as in GR).
This corresponds to the requirement that it should only
be possible to compare
lengths at one and the same world point. Second, it is
assumed, as in 
Riemannian geometry, that there is an affine 
(linear) torsion-free connection 
which defines a covariant derivative $\nabla$, and
respects the conformal structure.
Differentially this means that for any $g\in[g]$ the
covariant derivative $\nabla g$
should be proportional to $g$:
\be                                                \label{2.1}
\nabla g =-2A\otimes g\ \ \ \ \ \ \ 
(\nabla_{\lambda}g_{\mu\nu}=-2A_{\lambda}g_{\mu\nu}),
\ee
where $A=A_{\mu}dx^{\mu}$ is a differential 1-form.

Consider now a curve $\gamma: [0,1]\rightarrow M$ and a
parallel-transported
vector field $X$ along $\gamma$. If $l$ is the length of $X$,
measured with a representative $g\in[g]$, we obtain from (\ref{2.1})
the following relation between $l(p)$ for the initial point
$p=\gamma(0)$ and $l(q)$ for the end point $q=\gamma(1)$:
\be                                                \label{2.2}
l(q)=\exp\left(-\int_{\gamma}A\right)\ l(p).
\ee
Thus, the ratio of lengths in $q$ and $p$ (measured with
$g\in[g]$) {\it depends in general on the connecting path $\gamma$}
(see Fig.2). The length is only independent of $\gamma$ if the
curl of $A$,
\be                                                \label{2.3}
F=dA\ \ \ \  \ \ \ \ (F_{\mu\nu}=\partial_{\mu}A_{\nu}-
\partial_{\nu}A_{\mu}),
\ee
vanishes.
%\begin{center}
\begin{figure}
\begin{center}
\epsfig{file=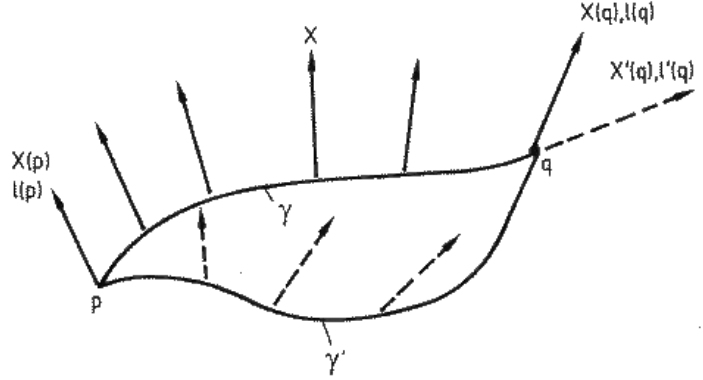,width=10cm}
\caption{Path dependence of parallel displacement and transport 
of length in  Weyl space.}
\end{center}
\end{figure}
%\clearpage

The compatibility requirement (\ref{2.1}) leads to the following
expression for the Christoffel symbols in Weyl's geometry:
\be                                                \label{2.4}
\Gamma^{\mu}_{\nu\lambda}=\frac{1}{2}g^{\mu\sigma}(
g_{\lambda\sigma,\nu}+g_{\sigma\nu,\lambda}-g_{\nu\lambda,\sigma})
+g^{\mu\sigma}(g_{\lambda\sigma}A_{\nu}+g_{\sigma\nu}A_{\lambda}-
g_{\nu\lambda}A_{\sigma}).
\ee
The second $A$-dependent term is a characteristic new piece
in Weyl's geometry which has to be added to the Christoffel symbols
of Riemannian geometry.

Until now we have chosen a fixed, but arbitrary metric in the
conformal class $[g]$. This corresponds to a choice of calibration
(or gauge). Passing to another calibration with metric $\bar{g}$,
related to $g$ by
\be                                                \label{2.5}
\bar{g}=e^{2\lambda}g,
\ee
the potential $A$ in (\ref{2.1}) will also change to $\bar{A}$, say.
Since the covariant derivative has an absolute meaning,
$\bar{A}$ can easily be worked out: On the one hand we have by
definition
\be                                                
\nabla \bar{g} =-2\bar{A}\otimes\bar{g}, 
\ee
and on the other hand we find for the left side with (\ref{2.1})
\be                                                %\label{2.7}
\nabla\bar{g}=\nabla(e^{2\lambda}g)=
2d\lambda\otimes\bar{g}+e^{2\lambda}\nabla g=
2d\lambda\otimes\bar{g}-2A\otimes\bar{g}.
\ee
Thus
\be                                                \label{2.6}
\bar{A}=A- d\lambda\ \ \ \ \ \
(\bar{A}_{\mu}=A_{\mu}-\partial_{\mu}\lambda).
\ee
This shows that a change of calibration of the metric induces a
{\it ``gauge transformation''} for $A$:
\be                                                \label{2.7}
g\rightarrow e^{2\lambda}g,\ \ \ \
A\rightarrow A-d\lambda.
\ee
Only gauge classes have an absolute meaning.
(The Weyl connection is, however, gauge-invariant.)

\subsection{Electromagnetism and Gravitation}

Turning to physics, Weyl assumes that his ``purely
infinitesimal geometry'' describes the structure of spacetime
and consequently he requires that physical laws should satisfy
a double-invariance: 1. They must be invariant with respect to
arbitrary smooth coordinate transformations.
2. They must be {\it gauge invariant}, i.e.,
invariant with respect to substitutions (\ref{2.7})
for an arbitrary smooth function $\lambda$.

Nothing is more natural to Weyl, than identifying $A_{\mu}$
with the vector potential and $F_{\mu\nu}$ in eq.(\ref{2.3})
with the field strength of electromagnetism.
In the absence of electromagnetic fields ($F_{\mu\nu}=0$)
the scale factor $\exp(-\int_{\gamma}A)$ in (\ref{2.2})
for length transport becomes path independent (integrable)
and one can find a gauge such that $A_{\mu}$ vanishes.
In this special case one is in the same situation as in GR.

Weyl proceeds to find an action which is generally invariant
as well as gauge invariant and which would give the coupled field
equations for $g$ and $A$. I do not want to enter into this,
except for the following remark. In his first paper \cite{2}
Weyl proposes what we call nowadays the Yang-Mills action
\be                                                   \label{2.8}
S(g,A)=-\frac{1}{4}\int Tr(\Omega\wedge\ast\Omega).
\ee
Here $\Omega$ denotes the curvature form and $\ast\Omega$
its Hodge dual\footnote{The integrand in (\ref{2.8}) is in
local coordinates indeed just the expression
$R_{\alpha\beta\gamma\delta} R^{\alpha\beta\gamma\delta}
\sqrt{-g}dx^{0}\wedge\ldots\wedge dx^{3}$ which is used
by Weyl ($R_{\alpha\beta\gamma\delta}$ $=$ curvature tensor of the
Weyl connection).}.
Note that the latter is gauge invariant, i.e., independent of the
choice of $g\in[g]$. In Weyl's geometry the curvature form
splits as $\Omega=\hat{\Omega}+F$, where $\hat{\Omega}$ is the
metric piece \cite{13}. Correspondingly, the action also splits,
\be                                                  \label{2.9}
Tr (\Omega\wedge\ast\Omega) =
Tr (\hat{\Omega}\wedge\ast\hat{\Omega})
+F\wedge\ast F.
\ee
The second term is just the Maxwell action. Weyl's theory
thus contains formally all aspects of a non-Abelian gauge theory.

Weyl emphasizes, of course, that the Einstein-Hilbert
action is not gauge invariant. Later work by Pauli \cite{14}
and by Weyl himself \cite{1,2} led soon to the conclusion that
the action (\ref{2.8}) could not be the correct one, and other
possibilities were investigated (see the later editions of RZM).

Independent of the precise form of the action Weyl shows that in
his theory gauge invariance implies the {\it conservation of electric
charge} in much the same way as general coordinate invariance
leads to the conservation of energy and momentum\footnote{I adopt here the somewhat naive interpretation of
energy-momentum conservation for generally invariant theories
of the older literature.}.
This beautiful connection pleased him particularly:
``\ldots [it] seems to me to be the strongest general argument
in favour of the present theory --- insofar as it is permissible
to talk of justification in the context of pure speculation.''
The invariance principles imply five `Bianchi type' identities.
Correspondingly, the five conservation laws follow in two
independent ways from the coupled field equations
and may be ``termed the eliminants'' of the latter. These structural connections
hold also in modern gauge theories.

\subsection{Einstein's objection and reactions of other physicists}

After this sketch of Weyl's theory I come to Einstein's
striking counterargument which he first communicated to Weyl
by postcard (see Fig.3). The problem is that if the idea of a
nonintegrable length connection (scale factor) is correct, 
then the behavior of clocks would depend on their history.
Consider two identical atomic clocks in adjacent world points
and bring them along different world trajectories which
meet again in adjacent world points.
According to (\ref{2.2}) their frequencies would
then generally differ. This is in clear contradiction with
empirical evidence, in particular with the existence of stable
atomic spectra. Einstein therefore concludes (see \cite{3}):
\begin{quotation}
%\item
{\sl \ldots (if) one drops the connection of the $ds$ to the
measurement of distance and time, then relativity looses all its
empirical basis.}
\end{quotation}

Nernst shared Einstein's objection and demanded on behalf of the
Berlin Academy that it should be printed in a short amendment to Weyl's
article, and Weyl had to cope with it. I have described the
intense and instructive subsequent correspondence between Weyl
and Einstein elsewhere \cite{3}. As an example, let me quote from
one of the last letters of Weyl to Einstein:
\begin{quotation}
%\item
{\sl This [insistence] irritates me of course, because
experience has proven that one can rely on your intuition;
so little convincing your counterarguments seem to me, as
I have to admit \ldots}
\end{quotation}

\begin{quotation}
%\item
{\sl By the way, you should not believe that I was driven to
introduce the linear differential form in addition to the
quadratic one by physical reasons. I wanted, just to the
contrary, to get rid of this `methodological inconsistency
{\it (Inkonsequenz)}' which has been a stone of contention
to me already much earlier. And then, to my surprise, I realized
that it looks as if it might explain electricity. You clap
your hands above your head and shout: But physics is not
made this way ! (Weyl to Einstein 10.12.1918).}
\end{quotation}

Weyl's reply to Einstein's criticism was, generally speaking, this:
The real behavior of measuring rods and clocks (atoms and atomic systems)
in arbitrary electromagnetic and gravitational fields can be
 deduced only from 
a dynamical theory of matter. 

Not all leading physicists reacted negatively. Einstein
transmitted a very positive first reaction by Planck, and
Sommerfeld wrote enthusiastically to Weyl that there was
``\ldots hardly doubt, that you are on the correct path and not
on the wrong one.''

In his encyclopedia article on relativity \cite{15} Pauli gave
a lucid and precise presentation of Weyl's theory, but
commented Weyl's point of view very critically. At the end he
states:
\begin{quotation}
%\item
{\sl \ldots Resuming one may say that Weyl's theory has not
yet contributed to get closer to the solution of the problem
of matter.}
\end{quotation}

Also Eddington's reaction was first very positive but he
changed his mind soon and denied the physical relevance of
Weyl's geometry.

The situation was later appropriately summarized by F.London
in his 1927 paper \cite{16} as follows:
\begin{quotation}
%\item
{\sl In the face of such elementary experimental evidence,
it must have been an unusually strong metaphysical conviction
that prevented Weyl from abandoning the idea that Nature would
have to make use of the beautiful geometrical possibility
that was offered. He stuck to his conviction and evaded
discussion of the above-mentioned contradictions through a rather
unclear re-interpretation of the concept of ``real state'',
which, however, robbed his theory of its immediate physical
meaning and attraction.}
\end{quotation}

\begin{figure}
\epsfig{file=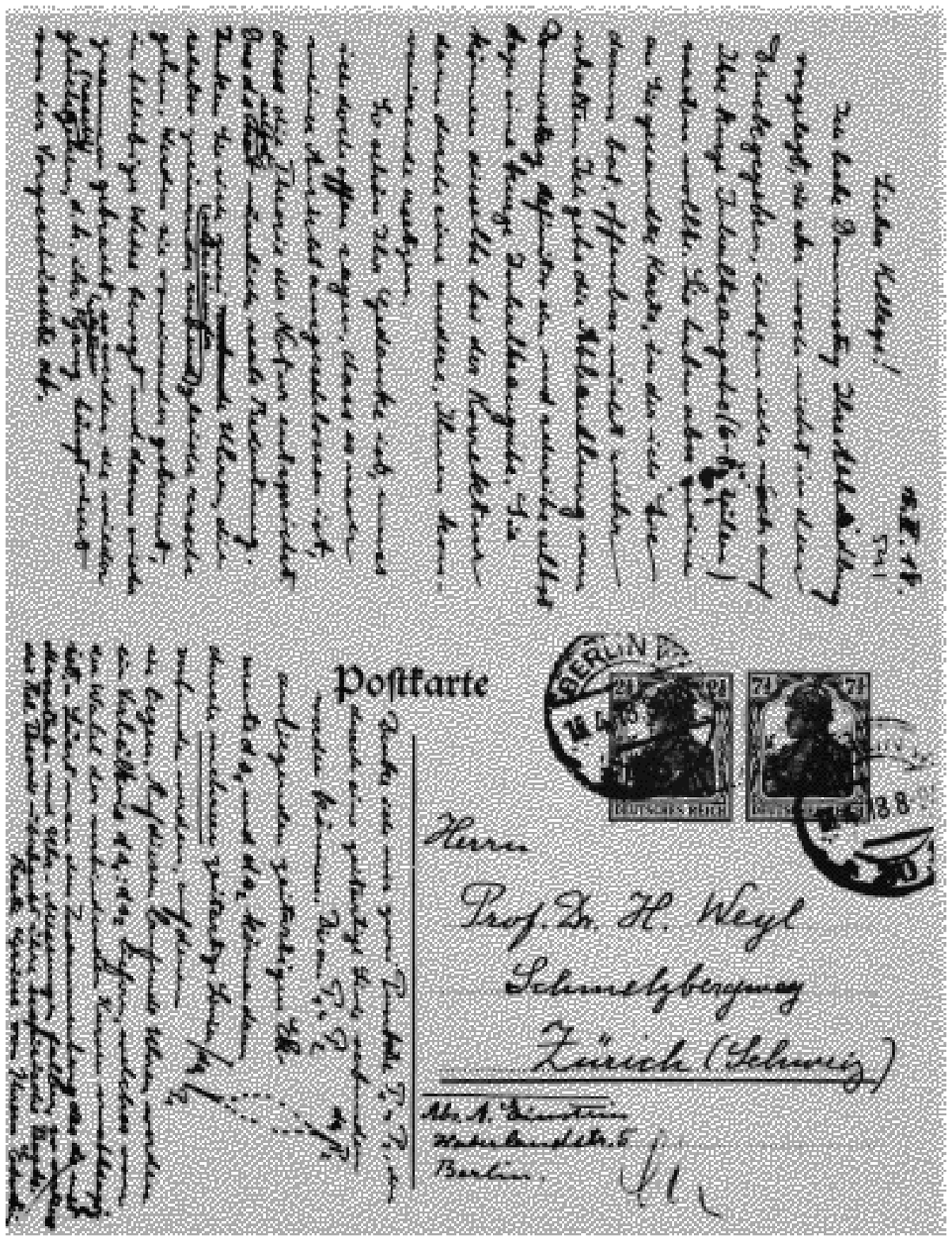,width=16cm}
\caption{Postcard of Einstein to Weyl 15.4.1918 (Archives of ETH).}
\end{figure}
\clearpage

\section{Weyl's 1929 Classic: ``Electron and Gravitation''}

Shortly before his death late in 1955, Weyl wrote for his
{\it Selecta} \cite{17} a postscript to his early attempt in
1918 to construct a `unified field theory'. There he expressed
his deep attachment to the gauge idea and adds (p.192):
\begin{quotation}
%\item
{\sl  Later the quantum-theory introduced the Schr\"odinger-Dirac
potential $\psi$ of the electron-positron field; it carried with it
an experimentally-based principle of gauge-invariance which
guaranteed the conservation of charge, and connected the $\psi$
with the electromagnetic potentials $\phi_{i}$ in the same way
that my speculative theory had connected the gravitational
potentials $g_{ik}$ with the $\phi_{i}$, and measured the
$\phi_{i}$ in known atomic, rather than unknown cosmological units.
I have no doubt but that the correct context for the principle
of gauge-invariance is here and not, as I believed in 1918, in the
intertwining of electromagnetism and gravity.}
\end{quotation}

This re-interpretation was developed by Weyl in one of the great
papers of this century \cite{4}. Weyl's classic does not only
give a very clear formulation of the gauge principle, but contains,
in addition, several other important concepts and results ---
in particular his two-component theory. The richness of the paper is
clearly visible from the following table of contents:
\begin{quotation}
%\item
{\sl Introduction. Relationship of General Relativity to the
quantum-theoretical field equations of the spinning electron:
mass, gauge-invariance, distant-parallelism. Expected modifications
of the Dirac theory. -I. Two-component theory: the wave function
$\psi$ has only two components. -$\S 1$. Connection between the
transformation of the $\psi$ and the transformation of a normal
tetrad  in four-dimensional space. Asymmetry of past and future,
of left and right. -$\S 2$. In General Relativity the metric
at a given point is determined by a normal tetrad. Components
of vectors relative to the tetrad and coordinates. Covariant
differentiation of $\psi$. -$\S 3$. Generally invariant form
of the Dirac action, characteristic for the wave-field of matter.
-$\S 4$. The differential conservation law of energy and momentum
and the symmetry of the energy-momentum tensor as a consequence
of the double-invariance (1) with respect to coordinate
transformations (2) with respect to rotation of the tetrad.
Momentum and moment of momentum for matter. -$\S 5$. Einstein's
classical theory of gravitation in the new analytic formulation.
Gravitational energy. -$\S 6$. The electromagnetic field. From
the arbitrariness of the gauge-factor in $\psi$ appears the necessity
of introducing the electromagnetic potential. Gauge invariance
and charge conservation. The space-integral of charge. The
introduction of mass. Discussion and rejection of another possibility
in which electromagnetism appears, not as an accompanying
phenomenon of matter, but of gravitation.}
\end{quotation}

The modern version of the gauge principle is already spelled out
in the introduction:

\begin{quotation}
%\item
{\sl The Dirac field-equations for $\psi$ together with the Maxwell
equations for the four potentials $f_{p}$ of the electromagnetic field
have an invariance property which is formally similar to the one which
I called gauge-invariance in my 1918 theory of gravitation and
electromagnetism; the equations remain invariant when one makes
the simultaneous substitutions
$$\psi\ \ \ {\rm by}\ \ \ e^{i\lambda}\psi\ \ \ \ {\rm and}\ \ \
f_{p}\ \ \ {\rm by}\ \ \  f_{p}-\frac{\partial\lambda}{\partial x^{p}},
$$
where $\lambda$ is understood to be an arbitrary function of position
in four-space. Here the factor $\frac{e}{ch}$, where $-e$ is the
charge of the electron, $c$ is the speed of light, and
$\frac{h}{2\pi}$ is the quantum of action, has been absorbed in
$f_{p}$. The connection of this ``gauge invariance'' to the
conservation of electric charge remains untouched.
But a fundamental difference, which is important to obtain agreement
with  observation, is that the exponent of the factor
multiplying $\psi$ is not real but pure imaginary. $\psi$ now
plays the role that Einstein's $ds$ played before. It seems to me
that this new principle of gauge-invariance, which follows not
from speculation but from experiment, tells us that the
electromagnetic field is a necessary accompanying phenomenon,
not of gravitation, but of the material wave-field represented
by $\psi$. Since gauge-invariance involves an arbitrary function
$\lambda$ it has the character of ``general'' relativity and can
naturally only be understood in that context.}
\end{quotation}

We shall soon enter into Weyl's justification which is, not
surprisingly, strongly associated with general relativity.
Before this I have to describe his incorporation of the Dirac theory
into GR which he achieved with the help of the tetrad formalism.

One of the reasons for adapting the Dirac theory of the spinning
electron to gravitation had to do with Einstein's recent
unified theory which invoked a distant parallelism with torsion.
E.Wigner \cite{18} and others had noticed a connection of this theory
and the spin theory of the electron. Weyl did not like this and
wanted to dispense with teleparallelism. In the introduction
he says:

\begin{quotation}
%\item
{\sl I prefer not to believe in distant parallelism for a number
of reasons. First my mathematical intuition objects to accepting
such an artificial geometry; I find it difficult to understand
the force that would keep the local tetrads at different points
and in rotated positions in a rigid relationship. There are,
I believe, two important physical reasons as well. The loosening
of the rigid relationship between the tetrads at different points
converts the gauge-factor $e^{i\lambda}$, which remains arbitrary
with respect to $\psi$, from a constant to an arbitrary function
of space-time. In other words, only through the loosening the
rigidity does the established gauge-invariance become
understandable. }
\end{quotation}

This thought is carried out in detail after Weyl has set up his
two-component theory in special relativity, including a discussion
of $P$ and $T$ invariance. He emphasizes thereby that the
two-component theory excludes a linear implementation of parity
and remarks: ``It is only the fact that the left-right symmetry
actually appears in Nature that forces us to introduce a second pair
of $\psi$-components.'' To Weyl the mass-problem is thus not
relevant for this. Indeed he says: ``Mass, however, is a
gravitational effect; thus there is hope of finding a substitute
in the theory of gravitation that would produce the required
corrections.''

We shall return to the two-component theory in $\S 5$ in
connection with parity violation and the $V-A$ interaction.

\subsection{Tetrad formalism}

The method of Weyl for incorporating his two-component spinors into
general relativity makes use of local tetrads (Vierbeins).

In the tetrad formalism the metric is described by an
arbitrary  basis of
orthonormal vector fields $\{e_{\alpha}(x);\alpha=0,1,2,3\}$.
If $\{e^{\alpha}(x)\}$ denotes the dual basis of 1-forms,
the metric is given by
\be                                                  \label{3.1}
g=\eta_{\mu\nu}e^{\nu}(x)\otimes e^{\nu}(x),\ \ \ \
(\eta_{\mu\nu})=diag(1,-1,-1,-1).
\ee
Weyl emphasizes, of course, that only a class of such local tetrads is
determined by the metric: the metric is not changed if the
tetrad fields are subject to spacetime-dependent Lorentz
transformations:
\be                                                  \label{3.2}
e^{\alpha}(x)\rightarrow\Lambda^{\alpha}_{\ \beta}(x)e^{\beta}(x).
\ee
With respect to a tetrad, the connection forms\footnote{I am using 
more modern notations;
for details see \cite{18}.}
$\omega=(\omega^{\alpha}_{\ \beta})$ have values in the Lie
algebra of the homogeneous Lorentz group:
\be                                                \label{3.3}
\omega_{\alpha\beta}+\omega_{\beta\alpha}=0.
\ee
(Indices are raised and lowered with $\eta^{\alpha\beta}$
and $\eta_{\alpha\beta}$, respectively.) They are determined
(in terms of the tetrad) by the first structure equation
of Cartan:
\be                                                \label{3.4}
de^{\alpha}+\omega^{\alpha}_{\ \beta}\wedge e^{\beta}=0.
\ee
Under local Lorentz transformations (\ref{3.2}) the connection
forms transform in the same way as the gauge potential of a non-Abelian gauge
theory:
\be                                                \label{3.5}
\omega(x)\rightarrow \Lambda(x)\omega(x)\Lambda^{-1}(x)-
d\Lambda(x)\Lambda^{-1}(x).
\ee
The curvature forms $\Omega=(\Omega^{\mu}_{\ \nu})$ are
obtained from $\omega$ in exactly the same way as the Yang-Mills field
strength from the gauge potential:
\be                                                \label{3.6}
\Omega=d\omega+\omega\wedge\omega
\ee
(second structure equation).

For a vector field $V$, with components $V^{\alpha}$ relative to
$\{e_{\alpha}\}$, the covariant derivative $DV$ is given by
\be                                                \label{3.7}
DV^{\alpha}=dV^{\alpha}+\omega^{\alpha}_{\ \beta}V^{\beta}.
\ee
Weyl generalizes this in a unique manner to spinor fields $\psi$:
\be                                                \label{3.8}
D\psi=d\psi+\frac{1}{4}\omega_{\alpha\beta}\sigma^{\alpha\beta}\psi.
\ee
Here, the $\sigma^{\alpha\beta}$ describe infinitesimal Lorentz
transformations (in the representation of $\psi$). For a Dirac
field these are the familiar matrices
\be                                               \label{3.9}
\sigma^{\alpha\beta}=\frac{1}{2}[\gamma^{\alpha},\gamma^{\beta}].
\ee
(For 2-component Weyl fields one has similar expressions in terms of
the Pauli matrices.)

With these tools the action principle for the coupled Einstein-Dirac
system can be set up. In the massless case the Lagrangian is
\be                                                   \label{3.10}
{\cal L}=\frac{1}{16\pi G}R-i\bar{\psi}\gamma^{\mu}D_{\mu}\psi,
\ee
where the first term is just the Einstein-Hilbert Lagrangian
(which is linear in $\Omega$). Weyl discusses, of course,
immediately the consequences of the following two symmetries:

(i) local Lorentz invariance,

(ii) general coordinate invariance.

\subsection{The new form of the gauge-principle}

All this is kind of a preparation for the final section of Weyl's paper,
which has the title ``electric field''. Weyl says:
\begin{quotation}
%\item
{\sl We come now to the critical part of the theory.
In my opinion the origin and necessity for the electromagnetic field
is in the following. The components $\psi_{1}$ $\psi_{2}$ are,
in fact, not uniquely determined by the tetrad but only to the
extent that they can still be multiplied by an arbitrary
``gauge-factor'' $e^{i\lambda}$. The transformation of the $\psi$
induced by a rotation of the tetrad is determined only up to such
a factor. In special relativity one must regard this gauge-factor
as a constant because here we have only a single
point-independent tetrad. Not so in general relativity;
every point has its own tetrad and hence its own arbitrary
gauge-factor; because by the removal of the rigid connection between
tetrads at different points  the gauge-factor necessarily
becomes an arbitrary function of position.}
\end{quotation}

In this manner Weyl arrives at the gauge-principle in its modern
 form and emphasizes:
``From the arbitrariness of the gauge-factor in $\psi$
appears the necessity of introducing the electromagnetic potential.''
The first term $d\psi$ in (\ref{3.8}) has now to be replaced
by the covariant gauge derivative $(d-ieA)\psi$ and the
nonintegrable scale factor (\ref{2.1}) of the old theory
is now replaced by a phase factor:
$$
\exp\left(-\int_{\gamma}A\right)\rightarrow
\exp\left(-i\int_{\gamma}A\right),
$$
which corresponds to the replacement of the original gauge
group {\bf R} by the compact group $U(1)$. Accordingly, the original
Gedankenexperiment of Einstein translates now to the
Aharonov-Bohm effect. The close connection between gauge
invariance and conservation of charge is again uncovered.
The current conservation follows, as in the original theory,
in two independent ways: On the one hand it is a consequence 
of the field equations for matter plus gauge invariance,
at the same time, however, also of the field equations
for the electromagnetic field plus gauge invariance.
This corresponds to an identity in the coupled system
of field equations which has to exist as a result of gauge invariance.
All this is nowadays familiar to students of physics and needs not
to be explained in more detail.

Much of Weyl's paper penetrated also into his classic book
``The Theory of Groups and Quantum Mechanics'' \cite{19}.
There he mentions also the transformation of his early
gauge-theoretic ideas: ``This principle of gauge invariance
is quite analogous to that previously set up by the author,
on speculative grounds, in order to arrive at a unified theory
of gravitation and electricity. But I now believe that this gauge
invariance does not tie together electricity and gravitation,
but rather electricity and matter.''

When Pauli saw the full version of Weyl's paper he became more
friendly and wrote \cite{20}:
\begin{quotation}
%\item
{\sl In contrast to the nasty things I said, the essential
part of my last letter has since been overtaken, particularly
by your paper in Z. f. Physik. For this reason I have afterward
even regretted that I wrote to you. After studying
your paper I believe
that I have really understood what you wanted to do
(this was not the case in respect of the little note in the
Proc.Nat.Acad.). First let me emphasize that side of the matter
concerning which I am in full agreement with you: your
incorporation of spinor theory into gravitational theory.
I am as dissatisfied as you are with distant parallelism and your
proposal to let the tetrads rotate independently
at different space-points is a true solution.}
\end{quotation}

In brackets Pauli adds:
\begin{quotation}
%\item
{\sl Here I must admit your ability in Physics.
Your earlier theory with $g'_{ik}=\lambda g_{ik}$ was pure
mathematics and unphysical. Einstein was justified in
criticizing and scolding. Now the hour of your revenge
has arrived.}
\end{quotation}

Then he remarks in connection with the mass-problem:
\begin{quotation}
%\item
{\sl Your method is valid even for the massive {\rm [Dirac]}
case. I thereby come to the other side of the matter, namely
the unsolved difficulties of the Dirac theory (two signs of
$m_{0}$) and the question of the 2-component theory.
In my opinion these problems will not be solved
by gravitation \ldots the gravitational effects will always be
much too small.}
\end{quotation}

Many years later, Weyl summarized this early tortuous
history of gauge theory in an instructive letter to the
Swiss writer and Einstein biographer C.Seelig, which
I reproduce in the German original \cite{21}.
\begin{quotation}
%\item
{\sl Aus dem Jahre 1918 datiert der von mir unternommene erste
Versuch, eine einheitliche Feldtheorie von Gravitation und
Elektromagnetismus zu entwickeln, und zwar auf Grund des
Prinzips der Eichinvarianz, das ich neben dasjenige der
Koordinaten-Invarianz stellte. Ich habe diese Theorie selber
l\"angst aufgegeben, nachdem ihr richtiger Kern: die
Eichinvarianz, in die Quantentheorie her\"uberge- rettet ist
als ein Prinzip, das nicht die Gravitation, sondern das
Wellenfeld des Elektrons mit dem elektromagnetischen verkn\"upft.
--- Einstein war von Anfang dagegen, und das gab zu mancher
Diskussion Anlass. Seinen konkreten Einw\"anden glaubte ich
begegnen zu k\"onen. Schliesslich sagte er dann:
``Na, Weyl, lassen wir das! So --- das heisst auf so
spekulative Weise, ohne ein leitendes, anschauliches
physikalisches Prinzip --- macht man keine Physik!''
Heute haben wir in dieser Hinsicht unsere Standpunkte wohl
vertauscht. Einstein glaubt, dass auf diesem Gebiet die Kluft
zwischen Idee und Erfahrung so gross ist, dass nur der Weg der
mathematischen Spekulation, deren Konsequenzen nat\"urlich
entwichelt und mit den Tatsachen konfrontiert werden m\"ussen,
Aussicht auf Erfolg hat, w\"ahrend mein Vertrauen in die reine
Spekulation gesunken ist und mir ein engerer Anschluss an die
quanten-physikalischen Erfahrungen geboten scheint, zumal es
nach meiner Ansicht nicht genug ist, Gravitation und
Elektromagnetismus zu einer Einheit zu verschmelzen.
Die Wellenfelder des Elektrons und was es sonst noch an
unreduzierbaren Elementarteilchen geben mag, m\"ussen mit
eigeschlossen werden.}
\end{quotation}

\section{Yang-Mills Theory}

In his Hermann Weyl Centenary Lecture at the ETH \cite{22},
C.N. Yang commented on Weyl's remark ``The principle of gauge-invariance
has the character of general relativity since it contains an arbitrary
function $\lambda$, and can certainly only be understood in terms
of it'' \cite{23} as follows:
\begin{quotation}
%\item
{\sl The quote above from Weyl's paper also contains something
which is very revealing, namely, his strong association of gauge
invariance with general relativity. That was, of course, natural since
the idea had originated in the first place with Weyl's attempt in 1918
to unify electromagnetism with gravity. Twenty years later, when
Mills and I worked on non-Abelian gauge fields, our motivation
was completely divorced from general relativity and we did not
appreciate that gauge fields and general relativity are somehow
related. Only in the late 1960's did I recognize the structural
similarity mathematically of non-Abelian gauge fields with general
relativity and understand that they both were connections
mathematically.}
\end{quotation}

Later, in connection with Weyl's strong emphasis of the relation
between gauge invariance and conservation of electric charge,
Yang continues with the following instructive remarks:
\begin{quotation}
%\item
{\sl Weyl's reason, it turns out, was also one of the melodies
of gauge theory that had very much appealed to me when as a
graduate student I studied field theory by reading Pauli's articles.
I made a number of unsuccessful attempts to generalize gauge theory
beyond electromagnetism, leading finally in 1954 to a
collaboration with Mills in which we developed a non-Abelian gauge
theory. In [\ldots ] we stated our motivation as follows:

The conservation of isotopic spin points to the existence of a
fundamental invariance law similar to the conservation of electric
charge. In the latter case, the electric charge serves as a source
of electromagnetic field; an important concept in this case is
gauge invariance which is closely connected with
(1) the equation of motion of the electro-magnetic field,
(2) the existence of a current density, and
(3) the possible interactions between a charged field and the
electromagnetic field. We have tried to generalize this concept
of gauge invariance to apply to isotopic spin conservation.
It turns out that a very natural generalization is possible.

Item (2) is the melody referred to above. The other two melodies,
(1) and (3), where what had become pressing in the early 1950's
when so many new particles had been discovered and physicists
had to understand now they interact which each other.

I had met Weyl in 1949 when I went to the Institute for Advanced
Study in Princeton as a young ``member''. I saw him from time
to time in the next years, 1949--1955. He was very approachable, 
but I don't remember having discussed physics or mathematics
with him at any time. His continued interest in the idea of gauge
fields was not known among the physicists. Neither
Oppenheimer nor Pauli ever mentioned it. I suspect they also 
did not tell Weyl of the 1954 papers of Mills' and mine.
Had they done that, or had Weyl somehow came across our paper,
I imagine he would have been pleased and excited, for we had
put together two things that were very close to his heart:
gauge invariance and non-Abelian Lie groups.}
\end{quotation}

It is indeed astonishing that during those late years Pauli
never talked with Weyl on non-Abelian generalizations of
gauge-invariance, since he himself had worked on this ---
even before the work of Yang and Mills. During a discussion
following a talk by Pais at the 1953  Lorentz Conference \cite{24}
in Leiden, Pauli said:
\begin{quotation}
%\item
{\sl \ldots I would like to ask in this connection whether the
transformation group [isospin] with constant phases can be
amplified in a way analogous to the gauge group for
electromagnetic potentials in such a way that the meson-nucleon
interaction is connected with the amplified group \ldots }
\end{quotation}

Stimulated by this discussion, Pauli worked on this problem
and drafted a manuscript to Pais that begins with \cite{8}:
\begin{quotation}
%\item
{\sl Written down July 22-25, 1953, in order to see how it looks.}
{\it Meson-Nucleon Interaction and Differential Geometry.}
\end{quotation}

Unaware of Klein's earlier contribution \cite{7}, Pauli generalizes
in this manuscript the Kaluza-Klein theory to a sixdimensional space,
and arrives through dimensional reduction at the essentials of an
$SU(2)$ gauge theory. The extra-dimensions are two-spheres with
spacetime dependent metrics on which $SU(2)$ operates in a
spacetime dependent manner. Pauli develops first in
``local language'' the geometry of what we now call a fiber bundle
with a homogeneous space as typical fiber (in his case
$S^{2}\cong SU(2)/U(1)$). Studying the curvature of the higher
dimensional space, Pauli automatically  finds for the first time
the correct expression for the non-Abelian field strengths.
Afterwards, Pauli sets up the 6-dimensional Dirac equation
and writes it out  in an explicit manner which is adapted to the
fibration. Later, in December 1953, he sends a ``Mathematical
Appendix'' to Pais and determines --- among other things ---
the mass spectrum implied by this equation. The final sentence reads:
``So this leads to some rather unphysical `shadow particles'.''
Pauli did not write down a Lagrangian for the gauge fields,
but as we shall see shortly, it was clear to him that the gauge
bosons had to be massless. This, beside the curious fermion
spectrum, must have been the reason why he did not publish anything.

With this background, the following story of spring 1954 becomes
more understandable. In late February, Yang was invited by
Oppenheimer to return to Princeton for a few days and to give
a seminar on his joint work with Mills. Here, Yang's report \cite{25}:
\begin{quotation}
%\item
{\sl  Pauli was spending the year in Princeton, and was deeply
interested in symmetries and interactions. (He had written in
German a rough outline of some thoughts, which he had sent to
A. Pais. Years later F.J. Dyson translated this outline into
English. It started with the remark, ``Written down July 22-25,
1953, in order to see how it looks,'' and had the title
``Meson-Nucleon Interaction and Differential Geometry.'')
Soon after my seminar began, when I had written down on the
blackboard,
$$
(\partial_{\mu}-i\epsilon B_{\mu})\psi,
$$
Pauli asked, ``What is the mass of this field $B_{\mu}$?''
I said we did not know. Then I resumed my presentation, but soon
Pauli asked the same question again. I said something to the effect
that that was a very complicated problem, we had worked on it
and had come to no definite conclusions. I still remember his
repartee: ``That is not sufficient excuse.'' I was so taken aback
that I decided, after a few moments' hesitation to sit down.
There was general embarrassment. Finally Oppenheimer said,
``We should let Frank proceed.'' I then resumed, and Pauli did not
ask any more questions during the seminar.

I don't remember what happened at the end of the seminar. But
the next day I found the following message:

February 24, Dear Yang, I regret that you made it almost impossible
for me to talk with you after the seminar. All good wishes.
Sincerely yours, W.Pauli.

I went to talk to Pauli. He said I should look up a paper by
E. Schr\"odinger, in which there were similar mathematics\footnote{E. Schr\"odinger, Sitzungsberichte der Preussischen
(Akademie der Wissenschaften, 1932), p. 105.}.
After I went back to Brookhaven, I looked for the paper and
finally obtained a copy. It was a discussion of
space-time-dependent representations of the $\gamma_{\mu}$
matrices for a Dirac electron in a gravitational field.
Equations in it were, on the one hand, related to equations
in Riemannian geometry and, on the other, similar to the
equations that Mills and I were working on. But it was many
years later when I understood that these were all different
cases of the mathematical theory of connections on fiber bundles.}
\end{quotation}

Later Yang adds:
\begin{quotation}
%\item
{\sl I often wondered what he [Pauli] would say about the subject
if he had lived into the sixties and seventies.}
\end{quotation}

At another occasion \cite{22} he remarked:
\begin{quotation}
%\item
{\sl I venture to say that if Weyl were to come back today,
he would find that amidst the very exciting, complicated and detailed
developments in both physics and mathematics, there are
fundamental things that he would feel very much at home with.
He had helped to create them.}
\end{quotation}

Having quoted earlier letters from Pauli to Weyl, I add what
Weyl said about Pauli in 1946 \cite{26}:
\begin{quotation}
%\item
{\sl The mathematicians feel near to Pauli since he is
distinguished among physicists by his highly developed organ for
mathematics. Even so, he is a physicist; for he has to a high
degree what makes the physicist; the genuine interest in the
experimental facts in all their puzzling complexity.
His accurate, instructive estimate of the relative weight
of relevant experimental facts has been an unfailing guide for
him in his theoretical investigations. Pauli combines in an
exemplary way physical insight and mathematical skill.}
\end{quotation}

To conclude this section, let me emphasize the main differences
of GR and Yang-Mills theories. Mathematically, the $so(1,3)$-valued
connection forms $\omega$ in $\S 3.1$ and the Liealgebra-valued
gauge potential $A$ are on the same footing; they are both
representatives of connections in (principle) fiber bundles
over the spacetime manifold. Eq.(\ref{3.6}) translates into
the formula for the Yang-Mills field strength $F$,
\be                                                  \label{4:1}
F=dA+A\wedge A.
\ee
In GR one has, however, additional geometrical structure, since the
connection derives from a metric, or the tetrad fields
$e^{\alpha}(x)$, through the first structure equation (\ref{3.4}).
Schematically, we have:

\begin{figure}
\begin{center}
\epsfig{file=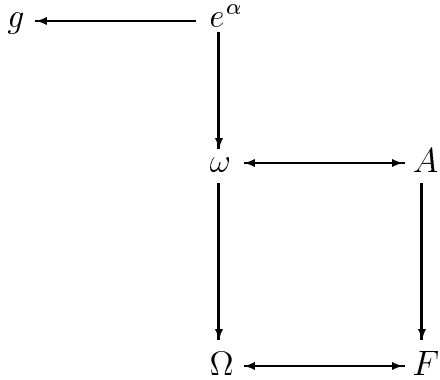,width=6cm}
\caption{General Relativity versus Yang-Mills theory.}
\end{center}
\end{figure}

(In bundle theoretical language one can express this as follows:
The principle bundle of GR, i.e., the orthonormal frame bundle,
is constructed from the base manifold and its metric, and has
therefore additional structure, implying in particular the
existence of a canonical 1-form (soldering form), whose
local representative are the tetrad fields; see, e.g. \cite{Bleecker}.)

Another important difference is that the gravitational Lagrangian
$\ast R=\frac{1}{2}\Omega_{\alpha\beta}\wedge\ast
(e^{\alpha}\wedge e^{\beta})$ is linear in the field strengths,
whereas the Yang-Mills Lagrangian $F\wedge\ast F$ is quadratic.

\section{Parity Violation and 2-Component Neutrino}

The two-component spinor theory was only briefly mentioned in my
discussion of Weyl's great 1929 paper. Since this massless spin $1/2$
equation became very important after the discovery of parity
violation I would now like to add  a few remarks.

Due to the fact that there exist two inequivalent irreducible
(projective) representations of the one-component of the
homogeneous Lorentz group, $L^{\uparrow}_{+}$ (with $SL(2,C)$ as
universal covering group), there are two types of fundamental
Weyl spinors, $\phi_{\alpha}$ and $\chi{\dot{\beta}}$, for which
the free Weyl equations read as follows:
\be                                                   \label{5.1}
\hat{\sigma}^{\mu}\partial_{\mu}\phi=0,\ \ \  \
\sigma^{\mu}\partial_{\mu}\chi=0.
\ee
Here, $(\sigma^{\mu})=(1 \! {\rm I},-\vec{\sigma})$,
$(\hat{\sigma}^{\mu})=(1 \!  {\rm I},\vec{\sigma})$ ($\vec{\sigma}$: Pauli
matrices). In spinor calculus these equations become
\be                                                 \label{5.2}
\partial^{\alpha\dot{\beta}}\phi_{\alpha}=0,\ \ \
\partial_{\alpha\dot{\beta}}\chi^{\dot{\beta}}=0.
\ee

In his ``New Testament'' from 1933 \cite{27}, Pauli rejected these
equations: ``Indessen sind diese Wellengleichungen, wie ja aus ihrer
Herleitung hervorgeht, nicht invariant gegen\"uber Spiegelungen
(Vertauschlung von links und recht) und infolgedessen sind sie auf die
physikalische Wirklichkeit nicht anwendbar.''

However, as long as no interactions are taken into account, this
statement is not correct. To make this evident one only has to note
that both equations in (\ref{5.1}) are equivalent to the
Majorana formulation: Consider, for instance, the $\phi$-field
and set
\be                                                    \label{5.3}
\psi=\left(\begin{array}{c}
\phi \\
\varepsilon\phi^{\ast}
\end{array}\right),\ \ \ \ \ \
\varepsilon=\left(\begin{array}{cc}
0 & 1 \\
-1 & 0
\end{array}\right),
\ee
then the first equation in (\ref{5.1}) is equivalent to the
massless Dirac equation,
\be                                                  \label{5.4}
\gamma^{\mu}\partial_{\mu}\psi=0.
\ee
Furthermore, $\psi$ is self-conjugate: A general Dirac spinor
$\left(\begin{array}{c}
\phi_{\alpha} \\
\chi^{\dot{\beta}}
\end{array}\right)$
transforms under charge conjugation $C$ according to
\be                                                 \label{5.5}
C:
\left(\begin{array}{c}
\phi \\
\chi
\end{array}\right)
\rightarrow
\left(\begin{array}{c}
-\varepsilon\chi^{\ast} \\
\varepsilon\phi^{\ast}
\end{array}\right)
\ee
and, for (\ref{5.3}), this reduces  to the Majorana condition
$C:\psi\rightarrow\psi$. Nobody would say that the Majorana theory
is not reflection invariant.

Note in this connection also the following: A Dirac field transforms
under $P$ as
\be                                               \label{5.6}
P:\psi\rightarrow\psi'(x)=\gamma^{0}\psi(Px).
\ee
For the Majorana field (\ref{5.3}) this translates into an
antilinear transformation for $\phi$,
\be                                                 \label{5.7}
P:\phi\rightarrow\phi'(x)=\varepsilon\phi^{\ast}(Px),
\ee
which leaves the Weyl equation invariant. Usually this
operation is interpreted as $CP$, but without interactions
this is a matter of semantics.

Before I will return to history, let me also remind you of the
formulation of Lee and Yang \cite{28}. These authors introduce
in the Weyl representation of the $\gamma$-matrices the Dirac
spinor
$\psi=\left(\begin{array}{c}
\phi \\
0
\end{array}\right)$,
whence $(1-\gamma^{5})\psi=0$. The first Weyl equation in (\ref{5.1})
is then again equivalent to the massless Dirac equation (\ref{5.4}).
In the Lee-Yang formulation one thus has
\be                                                \label{5.8}
\gamma^{\mu}\partial_{\mu}\psi=0,\ \ \ \ \
(1-\gamma^{5})\psi=0.
\ee
These equations are, of course, independent of the representation
of the $\gamma$-algebra.

Thus, the three formulations of Weyl, Majorana, and Lee-Yang are
entirely equivalent. This was noticed by several authors \cite{28}
shortly after the discovery of parity violation, but had been
worked out by J.Serpe \cite{29} already in 1952. Today, because
of the chiral nature of the fundamental fermions, the use of Weyl
spinors has become common practice.

The discovery of parity violation early in 1957 in several
experiments suggested by Lee and Yang \cite{30} was one of the
most exciting events in the fifties. Its impact was enormous,
as is illustrated by the following letter from Pauli to
Weisskopf \cite{31}:

\begin{quotation}
%\item
{\sl
\noindent
Dear Weisskopf,

Now the first shock is over and I begin to collect myself again (as one says
in Munich).

Yes, it was very dramatic. On Monday 21st at 8:15 p.m. I was supposed to give
a talk about ``past and recent history of the neutrino''. At 5 p.m. the mail
brought me three experimental papers: C.S. Wu, Lederman and Telegdi; the
latter
was so kind to send them to me. The same morning I received two theoretical
papers, one by Yang, Lee and Oehme, the second by Yang and Lee about the
two-component spinor theory. The latter was essentially identical
with the paper
by Salam, which I received as a preprint already six to eight weeks ago and 
to which I referred in my last short letter to you. (Was this paper known in
the USA?) ( At the same time came a letter from Geneva by Villars with the
New York Times article.)

Now, where shall I start? It is good that I did not make a bet. I would have
resulted in a heavy loss of money (which I cannot afford); I did make a fool
of myself, however (which I think I can afford to do)--- incidentally,
only in
letters or orally and not in anything that was printed. But the others now
have the right to laugh at me.

What shocks me is not the fact that ``God is just left-handed'' but the fact
that in spite of this He exhibits Himself as left/right symmetric when He
expresses Himself strongly. In short, the real problem now is why the strong
interaction are left/right symmetric. How can the strength of an interaction
produce or create symmetry groups, invariances or conservation laws? This
question prompted me to my premature and wrong prognosis. I don't know any
good answer to that question but one should consider that already there
exists
a precedent: the rotation group in isotopic spin-space,
which is not valid for
the electromagnetic field. One does not understand why it is valid at all.
It seems that there is a certain analogy here!

In my lecture I described how Bohr (Faraday lecture, 1932, Solvay Conference,
1932), as my main opponent in regard to the neutrino, considered plausible
the violation of the energy law in the beta-decay (what one calls today
``weak interaction''), how his opposition then became weaker and how he said
in a more general way (1933) that one must be ``prepared for surprises'' not 
anywhere but specifically with the beta-decay. Then I said spontaneously
(on the spur of the moment) that at the end of my talk I would come back
 to the surprises which Professor Bohr had foreseen here \ldots

Many questions, no answers !}
\end{quotation}

Let me say a bit more about the paper of Salam which is mentioned in Pauli's
letter. In September 1956 Salam had heard Yang's talk at the
Seattle Conference
on his and Lee's famous solution of the
$\vartheta -\tau$   puzzle by abandoning 
left/right symmetry in weak interactions. In his Nobel Price 
lecture Salam recollects \cite{32}:
\begin{quotation}
%\item
{\sl I remember travelling back to London on an American Air Force (MATS)
transport
flight. Although I had been granted, for that night, the status of a 
Brigadier
or a Field Marshal --- I don't quite remember which --- the plane was very 
uncomfortable, full of crying servicemen's children --- that is, the 
children were 
crying, not the servicemen. I could not sleep. I kept reflecting on why 
Nature should violate left/right symmetry in weak interactions. Now the 
hallmark of most weak interactions was the involvement in radioactivity
phenomena of Pauli's neutrino. While crossing over the Atlantic came back to
me a deeply perceptive question about the neutrino which Professor Rudolf
Peierls had asked when he was examining me for a Ph.D. a few years before.
Peierls' question was: ``The photon mass is zero because of Maxwell's
principle of a gauge symmetry for electromagnetism; tell me, why is the
neutrino mass zero?'' }
\end{quotation}

During that comfortless night he realized that Weyl's  two-component
equation for the neutrino would account for both parity violation and the 
masslessness of the neutrino. Soon afterwards he presented the idea to
Peierls, who replied: ``I do not believe left/right symmetry is violated
in weak forces at all.'' After that, Salam was hoping to find more resonance
at CERN. There he communicated the idea to Pauli, through Villars, who
``returned the next day with a message of the Oracle: Give my regards to my
friend Salam and tell him to think of something better.''

Meanwhile parity violation was discovered and Salam got a kind, apologetic
letter from Pauli. But this changed again soon afterwards. I quote:
\begin{quotation}
%\item
{\sl Thinking that Pauli's spirit should by now be suitably crushed,
I sent him
two short notes (Salam, 1957b) I had written in the meantime. These contained
suggestions to extend chiral symmetry to electrons and muons, assuming that
their masses were a consequence of what has come to be known as dynamical
spontaneous symmetry breaking. With chiral symmetry for electrons, muons,
and neutrinos, the only mesons that could mediate weak decays of the muons
would have to carry spin one. Reviving thus the notion of charged 
intermediate spin-one bosons, one could then postulate for these  a
type of gauge
invariance which I called the ``neutrino gauge''. Pauli's reaction was swift
and terrible. He wrote on 30 January 1957, then on 18 February and later
on 11, 12 and 13 March: ``I am reading (along the shores of Lake Zurich) in
bright sunshine quietly your paper \ldots'' ``I am very much startled on the
title of your paper `Universal Fermi Interaction' \ldots For quite a while
I have for myself the rule if a theoretician says universal it just means
pure nonsense. This holds particularly in connection with the Fermi 
interaction, but otherwise too, and now you too, Brutus, my son, come with 
this word \ldots'' Earlier, on 30 January, he had written:
``There is a similarity
between this type of gauge invariance and that which was published by Yang
and Mills... In the latter, of course, no $\gamma_{5}$ 
was used in the exponent,'' and
he gave me the full reference of Yang and Mills' paper
\cite{18}. I quote from
this letter:  ``However, there are dark points in your paper regarding the
vector field $B_{\mu}$.
If the rest mass is infinite (or very large), how can this
be compatible with the gauge transformation
$B_{\mu}\rightarrow B_{\mu}-\partial_{\mu}\Lambda $ ?'' and he concludes
his letter with the remark:  ``Every reader will realize that you deliberately
conceal here something and will ask you the same questions.}
\end{quotation}

\section{Chiral Invariance and Universal $V$--$A$ Interaction}

These recollections bring me to the last subject of my lecture. The   
two-component model of the neutrino paved also the way for a successful 
phenomenological description of weak interaction processes at low energies. 
In his masterly written review article ``On the earlier and
more recent history
of the Neutrino'' \cite{33}, Pauli remarks:
\begin{quotation}
%\item
{\sl For some time I faced this particular model with a
certain skepticism [42],
since it seemed to me that the special role of the neutrino was emphasized
too strongly. It turned out, however, that by further developing the ideas
of Stech and Jensen (see $\S 3$ above) the model allowed an
interesting generalization for the form of the interaction
energy for all weak interactions.}
\end{quotation}

After an inventory of the experimental situation, mentioning in particular
the new recoil experiments on ${\rm ^{6}He}$, Pauli continues with:
\begin{quotation}
%\item
{\sl Based on the Stech-Jensen transformation and the
two-component model of the
neutrino the following postulate suggests itself for the theoretical
interpretation:} {\it The Hamiltonian of each weak 4-fermion interaction shall
``universally'' contain either only R or only L components of the involved
fermions.} {\sl Equivalent to this postulate is the formulation that in the
transformation  $\psi'=\gamma_{5}\psi$ 
the density of the interaction energy for each
particle separately should ``universally'' 
remain unchanged or change its sign.}
\end{quotation}

At this point the classical papers \cite{34} are quoted, followed by the 
statement:
\begin{quotation}
%\item
{\sl The Stech-Jensen transformation referred to a pair of the particles
simultaneously while the two-component model of the neutrino is equivalent
to the validity of the result of the transformation for the neutrino alone.
The} {\it postulate of the extended Stech-Jensen transformation now under
discussion
is therefore a generalization of the two-component model of the neutrino.}
\end{quotation}

As we all know this postulate leads uniquely to the universal
$V$--$A$ interaction.
At the time it was disturbing that the $V$ and $A$ interaction strengths for
nucleons in beta decay are empirically not equal. Today we know that the
equality does hold on the level of the quark fields.

It is, unfortunately, not generally known that W. Theis 
proposed independently
the parity violating V-A interaction in a paper submitted on 20 December
1957 to the {\it Zeitschrift f\"ur Physik}
\cite{35}. Theis emphasized that in the spinor
calculus a Dirac spinor can be expressed in terms of a single two-component
Weyl spinor
\be                                                       \label{6.1}
\psi=\left(\begin{array}{c}
\phi_{\alpha} \\
\frac{i}{m}\partial^{\alpha\dot{\beta}}\phi_{\alpha}
\end{array}\right),
\ee
and that the Dirac equation is then equivalent to the Klein-Gordon equation
for $\phi_{\alpha}$. 
Since in this representation $\psi$
contains derivatives, the author
finds Fermi's requirement of a derivative-free coupling not so convincing
and requires instead a derivative-free four-Fermi interaction for the Weyl
spinors. This allows for only one possibility, namely
\be                                               \label{6.2}
p^{\ast}_{\alpha}n_{\dot{\beta}}e^{\ast\alpha}\nu^{\dot{\beta}}
+{\rm h.c.},
\ee
which is just the $V$--$A$ coupling.

This formal argument is similar to the one in the classic  paper  by 
Feynman
and Gell-Mann \cite{34}. The latter goes, however, beyond the
$V$--$A$ interaction
and advocates a current-current interaction Lagrangian, containing also
hypothetical self-terms. These imply processes like neutrino-electron
scattering or the annihilation process
$e^{-}+e^{+}\rightarrow\nu+\bar{\nu}$, which was soon
recognized to be very important in the later evolutionary stages of massive
stars \cite{36}. (We have heard a lot about this during the school.)

It may also not be known to the young generation that various experiments\footnote{For a description of the classic experiments, I refer to an
excellent paper by Telegdi \cite{37}.}
were in conflict with chiral invariance at the time when Feynman and
Gell-Mann wrote their paper. They had the courage to question the correctness
of these experiments:
\begin{quotation}
%\item
{\sl 
These theoretical arguments seem to the authors to be strong enough to
suggest that the disagreement with the
$^{6}He$ recoil experiment and with some
other less accurate experiments indicates that these experiments are wrong.
The $\pi\rightarrow e+\bar{\nu}$
problem may have a more subtle solution.}
\end{quotation}

The later verification of the prediction for the ratio $\Gamma(\pi\rightarrow e\nu)/
\Gamma(\pi\rightarrow\mu\nu)$ was one of the triumphs of the universal
$V$--$A$ interaction.

We will certainly hear more from J. Steinberger about the experimental side
of the story.

\section{Epilogue}

The developments after 1958 consisted in the gradual recognition
that --- contrary to phenomenological appearances --- Yang-Mills gauge theory
can describe weak and strong interactions. This important step was again
very difficult, with many hurdles to overcome.

One of them was the mass problem which was solved, perhaps in a preliminary
way, through spontaneous symmetry breaking. Of critical significance was
the recognition that spontaneously broken gauge theories are renormalizable.
On the experimental side the discovery and intensive investigation of the
neutral current was, of course, extremely crucial. For the gauge description of
the strong interactions, the discovery of asymptotic freedom was decisive .
That the $SU(3)$ color group should be gauged was also not at all obvious.
And then there was the confinement idea which explains why quarks and gluons
do not exist as free particles. All this is described in numerous modern text
books and does not have to be repeated.

The next step of creating a more unified theory of the basic interactions
will probably be much more difficult. All major theoretical
developments of the last twenty years, such as grand unification,
supergravity
and supersymmetric string theory are almost completely separated from
experience. There is a great danger that theoreticians get lost in pure
speculations. Like in the first unification proposal of Hermann Weyl they may
create beautiful and highly relevant mathematics which does, however, not
describe nature. Remember what Weyl wrote to C. Seelig in his late years:
\begin{quotation}
%\item
{\sl Einstein glaubt, dass auf diesem Gebiet die Kluft zwischen 
 Idee und
Erfahrung so gross ist, dass nur der Weg der mathematischen Spekulation
(\ldots)
Aussicht auf Erfolg hat, w\"ahrend mein Vertrauen in die reine Spekulation
gesunken ist \ldots }
\end{quotation}

%\newpage


\begin{thebibliography}{99}

\bibitem{1} H. Weyl, {\it Space $\cdot$ Time $\cdot$ Matter.} Translated 
from the 4th German Edition. London: Methmen 1922. 
{\it Raum $\cdot$ Zeit $\cdot$ Materie}, 8.
Auflage, Springer-Verlag (1993).

\bibitem{2} H. Weyl, {\it Gravitation und Elektrizit\"at.}
Sitzungsberichte Akademie der Wissenschaften Berlin, 465-480 (1918).
Siehe auch die {\em Gesammelten Abhandlungen.} 6
Vols.\ Ed.\ K.\ Chadrasekharan, Springer-Verlag.

\bibitem{3} N. Straumann, {\em Zum Ursprung der Eichtheorien bei
Hermann Weyl.} Physikalische Bl\"atter {\bf 43} (11), 414-421 (1987).

\bibitem{4} H. Weyl, {\it Elektron und Gravitation. I.} Z.\ Phys.\ {\bf
56}, 330 (1929).

\bibitem{5} W.Pauli, {\em Wissenschaftlicher Briefwechsel}, Vol.\ I:
1919-1929. p.\ 505. Springer-Verlag 1979. (Translation of the letter by
L.\ O'Raifeartaigh.)

\bibitem{6} L. O'Raifertaigh, {\it The Dawning of Gauge Theory.}
Princeton University Press, to appear.

\bibitem{7} O. Klein, {\it On the Theory of charged Fields.} 1938
Conference on New Theories in Physics held at Kazimierz, Poland 1938.

\bibitem{8} W. Pauli, {\it Meson-Nucleon Interaction and Differential
Geometry.} Letter to Pais, to appear in [6].

\bibitem{9} R. Shaw, Thesis, Cambridge University 1955.

\bibitem{10} R. Utiyama, {\it Butsurigaku wa dokumade susunkada} (How
far has Physics progressed). Iwanami Shoten, Tokyo 1983.

\bibitem{11} N. Straumann. {\em Urspr\"unge der Eichtheorien.}
DMV-Seminar Geschichte der Mathematik (H.\ Weyls ``Raum-Zeit-Materie''),
to appear in Springer-Verlag (1996).

\bibitem{12} N. Straumann {\it Von der Stech-Jensen-Transformation zur
universellen V-A Wechselwirkung}.  Archive for History of Exact
Sciences, {\bf 44}, 365 (1992).

\bibitem{13} J. Audretsch, F.G\"ahler and N. Straumann, {\it
Comm.Math.Phys.} {\bf 95}, 41 (1984).

\bibitem{14} W. Pauli, {\it Zur Theorie der Gravitation und der
Elektrizit\"at von H. Weyl.} Physikalische Zeitschrift {\bf 20},
457-467 (1919).

\bibitem{15} W. Pauli, {\it Relativit\"atstheorie.} Encyklop\"adie der
Mathematischen Wissenschaften 5.2, Leipzig:  Teubner, 539-775 (1921).

\bibitem{16} F. London, {\it Quantenmechanische Deutung der Theorie von
Weyl.} Z.\ Phys.\ {\bf 42}, 375 (1927).

\bibitem{17} H. Weyl, {\it Selecta.} Birkh\"auser-Verlag 1956.

\bibitem{18} N. Straumann, {\it General Relativity and Relativistic
Astrophysics.}  Texts and Monographs in Physics, Springer-Verlag
(1984).

\bibitem{19} H. Weyl, {\it Gruppentheorie und Quantenmechanik.}
Wissenschaftliche Buchgesellschaft, Darmstadt 1981 (Nachdruck der
2.\ Aufl., Leipzig 1931). Engl. translation: ``Group Theory and Quantum
Mechanics'', Dover, New York, 1950.

\bibitem{20} Ref.[5], p. 518.

\bibitem{21} In Carl Seelig: {\it Albert Einstein}.  Europa Verlag
Z\"urich 1960, p. 274.

\bibitem{22} C.N. Yang, {\it Hermann Weyl's Contribution to Physics.}
In: {\it Hermann Weyl}, Edited by K. Chandrasekharan, Springer-Verlag
1980.

\bibitem{23} H. Weyl, {\it Gesammelte Abhandlungen}, Vol.I to IV.
Springer-Verlag 1968. Edited by K. Chandrasekharan.  (Vol.III, p.229).

\bibitem{24} Conference in Honour of H.A. Lorentz, Leiden 1953.  {\it
Proceedings in Physica}, {\bf 19} (1953). A.Pais, p. 869.

\bibitem{25} C.N. Yang, {\it Selected Papers 1945-1980 with
Commentary.} Freeman and Co. 1983, p.525.

\bibitem{26}  H. Weyl. {\em Memorabilia.} Ref.\ \cite{22}, p. 85.

\bibitem{27} W. Pauli, {\it Die allgemeinen Prinzipien der
Wellenmechanik.} Handbuch der Physik, Geiger und Scheel, 2. Auff.,
Vol.24, Teil 1 (1933).

Von N. Straumann neu herausgegeben und mit historischen Anmerkungen
versehen, Springer-Verlag (1990). English translation: {\it General
Principles of Quantum Mechanics.} Springer-Verlag 1980.

\bibitem{28} T.D. Lee, C.N. Yang, {\it Phys.Rev.} {\bf 105}, 1671
(1957);

L.D. Landau, {\it Nucl.Phys.} {\bf 3}, 127 (1957);

A. Salam, {\it Nuovo Cimento} {\bf 5}, 299 (1957);

K.M. Case, {\it Phys.Rev.} {\bf 107}, 307 (1957);

J.A. McLennan, Jr. {\it Phys.Rev.} {\bf 106}, 821 (1957).

\bibitem{29} J. Serpe, {\it Physica} {\bf 18}, 295 (1952).

\bibitem{30} T.D. Lee and C.N. Yang, {\it Phys.Rev.} {\bf 104}, 254
(1956).

\bibitem{31} In Ref.[5]. Translation by V.F. Weisskopf.

\bibitem{32} A. Salam, {\it Gauge Unification and fundamental Forces},
{\it Rev.Mod.Phys.} {\bf 52}, 525 (1980).

\bibitem{33} W. Pauli, {\em Zur \"alteren und neueren Geschichte des
Neutrinos,} in W. Pauli: Aufs\"atze und Vortr\"age \"uber Physik und
Erkenntnistherorie. Vieweg, Braunschweig 1961. English translation in:
{\em Neutrino Physics,} Ed.: K. Winter, Cambridge University Press 1991,
p.1.

\bibitem{34} E.C.G. Sudarshan and R.E. Marshak, {\it Phys.Rev.} {\bf
19}, 1860 (1958);

J.J. Sakurai, {\it Nuovo Cimento} {\bf 7}, 649 (1958);

R.P. Feynman and M. Gell-Mann, {\it Phys.Rev.} {\bf 109}, 193 (1958).

\bibitem{35} W.R. Theis, {\it Z.Physik} {\bf 150}, 590 (1958).

\bibitem{36} H.Y. Chiu, P. Morrison, {\it Phys.Rev.Lett.} {\bf 5}, 573
(1960).

\bibitem{37} V.L.Telegdi, {\it The early experiments leading to V-A
interaction.} In: {\it ``Pions to Quarks: particle physics in the
1950's''}, Ed: L.M. Brown, M. Dresden, Cambridge University Press
1989.

\bibitem{Bleecker} D. Bleecker, {\it Gauge Theory and Variational
Principles}, Addison-Wesley 1981.




\end{thebibliography}
\end{document}